\documentclass[aip,apl,twocolumn,superscriptaddress]{revtex4}
\usepackage{epsfig}
\usepackage[usenames]{color}
\DeclareMathAlphabet{\mathitb}{OT1}{cmr}{bx}{sl}

\begin{document}

\renewcommand{\thefootnote}{\fnsymbol{footnote}}

\title{Back-action Induced Non-equilibrium Effect in Electron Charge Counting Statistics}

\author{HaiOu Li}
\author{Ming Xiao}
\email{maaxiao@ustc.edu.cn}
\author{Gang Cao}
\author{Cheng Zhou}
\author{RuNan Shang}
\author{Tao Tu}
\author{GuangCan Guo}
\affiliation{Key Laboratory of Quantum Information, Chinese Academy of Sciences, University of Science and Technology of China, Hefei 230026, People's Republic of China}
\author{HongWen Jiang}
\affiliation{Department of Physics and Astronomy, University of California at Los Angeles, 405 Hilgard Avenue, Los Angeles, CA 90095, USA}
\author{GuoPing Guo}
\email{gpguo@ustc.edu.cn}
\affiliation{Key Laboratory of Quantum Information, Chinese Academy of Sciences, University of Science and Technology of China, Hefei 230026, People's Republic of China}

\date{\today}

\begin{abstract}
We report our study of the real-time charge counting statistics measured by a quantum point contact (QPC) coupled to a single quantum dot (QD) under different back-action strength. By tuning the QD-QPC coupling or QPC bias, we controlled the QPC back-action which drives the QD electrons out of thermal equilibrium. The random telegraph signal (RTS) statistics showed strong and tunable non-thermal-equilibrium saturation effect, which can be quantitatively characterized as a back-action induced tunneling out rate. We found that the QD-QPC coupling and QPC bias voltage played different roles on the back-action strength and cut-off energy.
\end{abstract}

\maketitle

A quantum point contact (QPC) nearby a quantum dot (QD) is widely used to perform electron charge counting, which is important in many aspects such as the read-out of electron charge- or spin-based qubits \cite{Petta-Charge-Qubit-DQD, Delft-Single-Spin-Readout}. However, the measurement with a QPC has inevitable side effects, known as the back-action \cite{Kouwenhoven-Backaction-Theory}. For example, the back-action can drive the QD electrons out of thermal equilibrium, facilitating ground to excited-state transitions \cite{Ensslin-Backaction-eVqpc, Ludwig-Backaction-QPC-Current}, and possibly causing qubit relaxation and dephasing \cite{Weizman-Backaction-WhichPath}. Thus, cautions have to be exercised to minimize the back-action, in order to use the QPC as a non-invasive read-out sensor.  

However, back-action is usually indirect and hard to quantitatively analyze. In QDs the excited tunneling through higher-energy levels has been utilized to study back-action\cite{Kouwenhoven-Backaction-Theory, Ensslin-Backaction-eVqpc, Ludwig-Backaction-QPC-Current, Gustavsson-Backaction-Terahertz}. Here we found that we can relate the non-thermal-equilibrium effect in the charge counting statistics to back-action. This effect arises from the back-action induced tunneling directly out of the QD, without the help of excited levels. We developed a phenomenological model to quantitatively characterize the back-action strength. We also found the different contributions of the QD-QPC Coulomb coupling and QPC bias voltage. Especially we revealed that the back-action cut-off energy is proportional to the QD-QPC coupling, whereas independent of the QPC bias.    

\begin{figure}[t]
\begin{center}
\epsfig{file=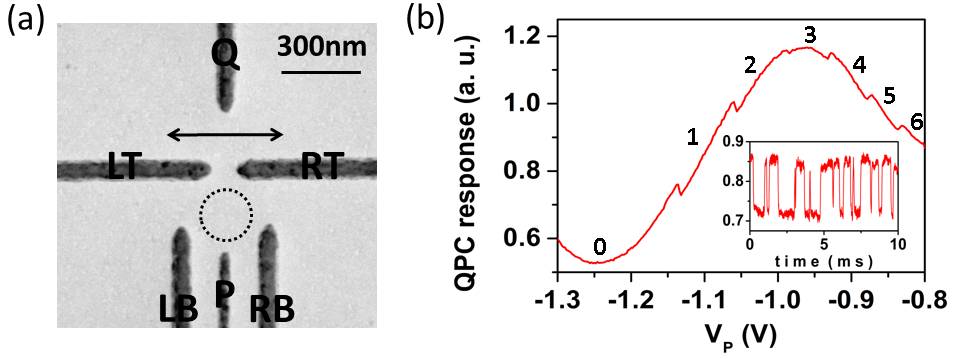, width=1\linewidth, angle=0}
\end{center}
\vspace{-7.5mm}
\caption{(a) A SEM picture of the surface gates. The open circle is where the QD is defined. The arrowed line shows the QPC channel. (b) A charge counting trace when $V_{P}$ is swept to squeeze the QD electron number from 6 to 1. The insert shows a real-time charge counting trace with fixed $V_{P}=-1.11V$, corresponding to the $0e \leftrightarrow 1e$ transition.}
\label{Figure1}
\end{figure}

We fabricated a QD with a QPC on side in a GaAs/AlGaAs heterostructure. The two-dimensional electron gas (2DEG) is 95 nm below the surface. The 2DEG has a density of $3.2\times10^{11} cm^{−2}$ and a mobility of $1.5\times10^{5} cm^{2}V^{-1}s^{-1}$. Fig. \ref{Figure1} (a) shows the scanning electron microscopy (SEM) picture of the surface gates. Five gates LT, RT, LB, RB and P shaped the QD. Gate Q, along with LT and RT, formed a QPC channel to count the QD electron number via capacitive coupling. A small gap between LT and RT was created to maximize this coupling. The experiment was done in a Helium-3 refrigerator with base temperature of $240 mK$. We operate the QD in such a way that the left barrier is closed and the electrons only tunnel through the right barrier (tunneling rate conveniently controlled by gate RB). The voltages on LT and RT are set below pinch-off, so that no leakage tunneling to the QPC was found. Both the source and drain of the QD are grounded.  A small dc bias is applied through the QPC channel. Fig. \ref{Figure1} (b) shows the QPC response while gate P is used to control the number of QD electrons. The insert is a trace of random telegraph signal (RTS) for the $0e \leftrightarrow 1e$ transition.

\begin{figure}[t]
\begin{center}
\epsfig{file=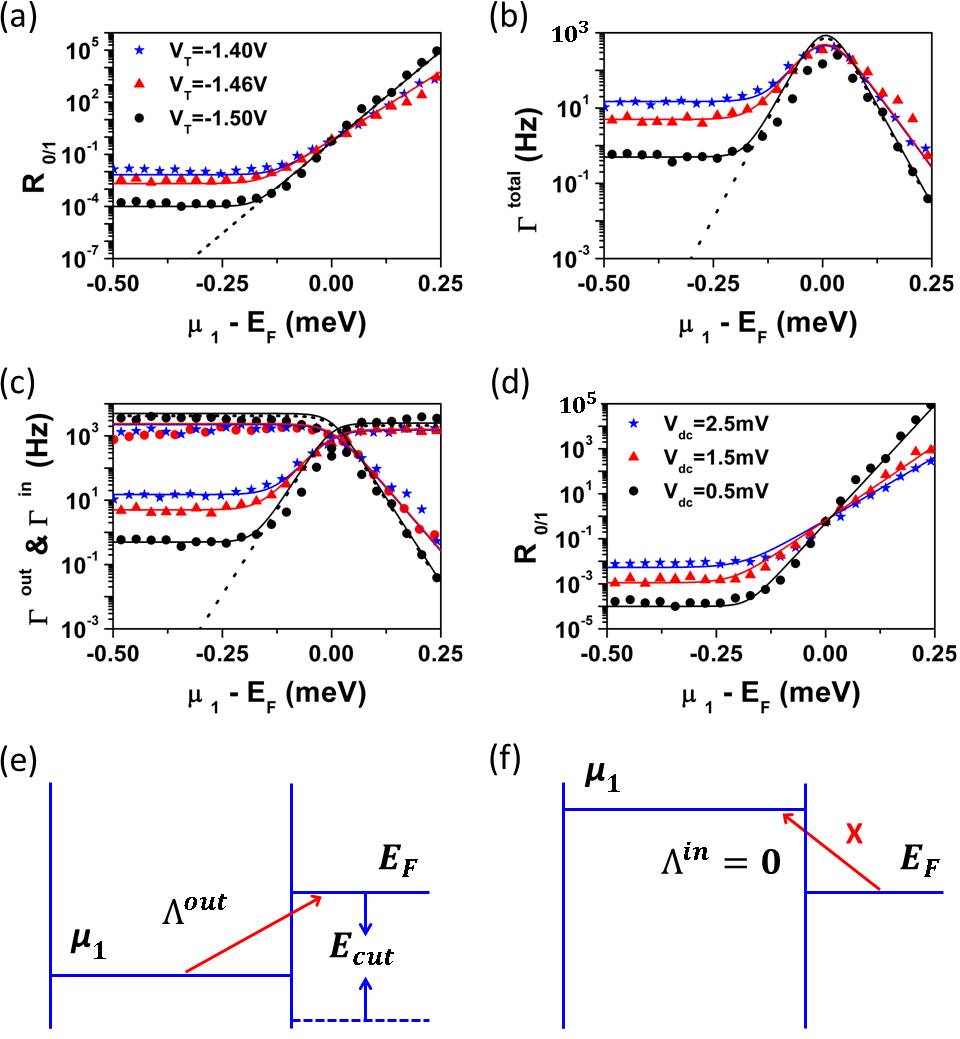, width=1\linewidth, angle=0}
\end{center}
\vspace{-7.5mm}
\caption{((a) - (c) RTS statistics ($R_{0/1}$, $\Gamma^{total}$, and $\Gamma^{out} \& \Gamma^{in}$), for the $0e \leftrightarrow 1e$ transition when $V_{T}$ is varied and $V_{dc}$ is fixed at $0.5mV$. For black circles, red triangles, and blue stars, $V_{T}= -1.50V$, $-1.46V$, and $-1.40V$, respectively. The dotted lines are simulation for a perfect thermal-equilibrium RTS. The solid lines are simulation of our own back-action model. (d) $R_{0/1}$ at fixed $V_{T}=-1.50V$ and varying $V_{dc}$. The black circles, red triangles, and blue stars corresponds to $V_{dc}$ from $0.5mV$ through $1.5mV$ to $2.5mV$, respectively. (e) Illustration of back-action assisted tunneling. When $0 < E_{F} - \mu_{1} < E_{cut}$, the QD electron absorbs phonon energy and tunnels out of the dot. Here $E_{cut}$ means the maximum phonon energy that the QD electron absorbs. (f) When $E_{F} - \mu_{1} > 0$, no obvious effect since 2DEG is a huge reservoir in thermal equilibrium. }
\label{Figure2}
\end{figure}

When studying the RTS statistics carefully, in some conditions we found large deviation from a thermal equilibrium picture. We studied all the last six electrons and found similar phenomena, except the complication of additional tunneling channels through spin excited levels for and only for the even electron numbers \cite{Ming-Backaction-SpinST}. Here we focus on the $0e \leftrightarrow 1e$ transition which contains all the basic features without this complication . Fig. \ref{Figure2} (a) - (c) showed the RTS statistics with different QD-QPC gap opening (controlled by voltage $V_{T} \equiv  V_{LT}=V_{RT}$), including the $0e$ and $1e$ occupancy ratio $R_{0/1}$,  total tunneling rate $\Gamma^{total}$, tunneling out rate $\Gamma^{out}$, and tunneling in rate $\Gamma^{in}$. In comparison we also showed the simulation of the RTS statistics for a thermally activated two-level switching in dotted lines \cite{Uren-RTS-Review}:
\begin{center}
$\left\{ \begin{array}{llll}
\Gamma^{out}=g_{n-1} \Gamma^{*}  (1-f(\mu_{n})) \\
\Gamma^{in}= g_{n} \Gamma^{*} f(\mu_{n})  \\
\Gamma^{total}=1/(1/\Gamma^{out}+1/ \Gamma^{in})  \\
R_{n-1/n}=\Gamma^{out}/\Gamma^{in}=g_{n-1}/g_{n}e^{(\mu_{n}-E_{F}) / k_{B}T}  \\
\end{array} \right. $
\end{center}

Here $g_{n}$ is the spin degeneracy; $\mu_{n}$ denotes the addition energy for the $n^{th}$ electron; Maximum tunneling rate $\Gamma^{*} \equiv (2\pi / \hbar) D \Delta^{2}$ where $D$  the electron density of energy and $\Delta$ is the tunneling matrix element; $f(\mu_{n})$ is the Fermi distribution function. We used the energy-voltage conversion factor $0.086 meV/mV$ obtained in transport experiment and temperature 240mK read by a sensor. 

Our data can be well described by the thermal equilibrium equations when $\mu_{n}-E_{F}>0$. However, a deviation appears when $\mu_{n}-E_{F}<0$, where the first electron is mostly trapped in the QD. $R_{0/1}$ should exponentially decay when $\mu_{1}$ drops below $E_{F}$ since the electron loses energy to tunnel out of the QD. Not as expected, we observed that $R_{0/1}$ saturates in this region. The same phenomena happens for the tunneling rates $\Gamma^{total}$ and $\Gamma^{out}$, but not for $\Gamma^{in}$. This saturation effect is found to increase with either the QD-QPC coupling or the QPC dc bias voltage.  For example, in Fig. \ref{Figure2} (a) we see that when $V_{T}=-1.40V$, $R_{0/1}$ starts to saturate when its value drops to $10^{-2}$, two magnitude larger than the saturation point $10^{-4}$ under the $V_{T}=-1.50V$ condition. We also observed stronger saturation effect by increasing the QPC dc bias voltage $V_{dc}$, as shown in Fig. \ref{Figure2} (d).

Although not open enough to allow direct tunneling, the QD-QPC gap increases their coupling strength drastically. The RTS amplitude reaches as high as $20 \%$. Such huge coupling must mean strong back-action as well. Principally the back-action strength is determined by the QPC condition and QD-QPC interaction together. The former can be controlled by the QPC dc bias voltage $V_{dc}$ \cite{Ensslin-Backaction-eVqpc, Ludwig-Backaction-QPC-Current}, and the latter can be controlled by voltage $V_{T}$. So by increasing either $V_{dc}$ or $V_{T}$ we introduce stronger back-action. The effect of back-action is illustrated in Fig. \ref{Figure2} (e) and (f). When $\mu_{1}$ is below $E_{F}$, the QD should have been mostly filled with one electron. However, after receiving phonons emitted by the QPC, the electron gains energy and tunnels out. The outcome is that the QD empty occupancy increases and so does the ratio $R_{0/1}$. This phonon-assisted tunneling remains there when $E_{F} - \mu_{1}$ is less than a certain cut-off energy \cite{Ludwig-Backaction-DQD}. On the other hand, when $\mu_{1}$ is way above $E_{F}$, the phonon doesn't apparently assist the electrons to tunnel inside since the 2DEG is supposed to be a huge reservoir in thermal equilibrium. This explains why $\Gamma^{in}$ doesn't show saturation effect. We developed a phenomenological model about the effect of back-action:
\begin{center}
$\left\{ \begin{array}{llll}
\Gamma^{out}=g_{n-1} [\Lambda^{out} + \Gamma^{*}  (1-f(\mu_{n}))] \\
\Gamma^{in}= g_{n}  [\Lambda^{in} + \Gamma^{*} f(\mu_{n})] \\
\Gamma^{total}=1/(1/\Gamma^{out}+1/ \Gamma^{in}) \\
R_{n-1/n}=\Gamma^{out}/\Gamma^{in} \\
\end{array} \right. $
\end{center}

Here we introduced two extra tunneling rates $\Lambda^{out}$ and $\Lambda^{in}$. They refer to the back-action driven tunneling out and in rates. Their values can be easily estimated from $\Gamma^{out}$ and $\Gamma^{in}$ in the extreme conditions:  $\Gamma^{out} \approx g_{n-1}\Lambda^{out}$, $\Gamma^{in} \approx  g_{n} \Gamma^{*}$ when $\mu_{n}<<E_{F}$; and $\Gamma^{out} \approx g_{n-1} \Gamma^{*}$, $\Gamma^{in} \approx g_{n}\Lambda^{in}$ when $\mu_{n}>>E_{F}$. We can immediately tell two basic features: $\Lambda^{out}$ is invariant and $\Lambda^{in}$ is negligible. $\Lambda^{out}$ is constant since the saturation tails are flat, implying a constant phonon spectrum before a certain cut-off energy. $\Lambda^{in}$ is found to be at least two-magnitude less than $\Lambda^{out}$ and two more magnitudes smaller than the other tunneling rates so we can safely ignore it. We showed our simulation as solid lines in Fig. \ref{Figure2} (a) -(d). Good agreement with the experiment was found.

\begin{figure}[t]
\begin{center}
\epsfig{file=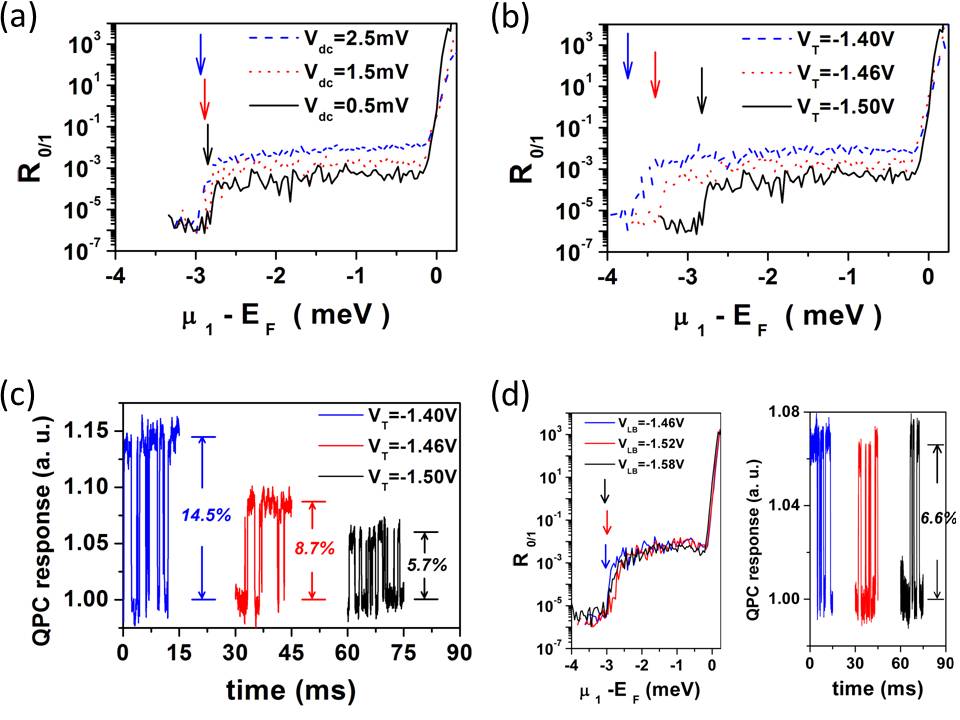, width=1\linewidth, angle=0}
\end{center}
\vspace{-7.5mm}
\caption{(a) $R_{0/1}$ versus $V_{dc}$ at fixed $V_{T}=-1.50V$. Arrows indicate the cut-off energy. (b) $R_{0/1}$ versus $V_{T}$ at fixed $V_{dc}=0.5mV$.  (c) RTS amplitude expressed as the percentage change under different $V_{T}$. All the data are normalized for comparison. (d) Both $R_{0/1}$ and the RTS amplitude show no change for varying $V_{LB}$ when $V_{T}$ is fixed at $-1.50V$.}
\label{Figure3}
\end{figure}

We summarized all the simulation results in table I. As $V_{T}$ or $V_{dc}$ is increased, the back-action induced tunneling out rate $\Lambda^{out}$ shows observable increase. We use the percentage of $\Lambda^{out}$ on $\Gamma^{*}$ as a more objective measure since $\Gamma^{*}$ or $\Lambda^{out}$ alone could be affected in different experimental conditions. Table I shows that $\Lambda^{out}/ \Gamma^{*}$ steadily increases from $0.026\%$ to $1\%$ as we increase $V_{T}$ alone, and from $0.026\%$ to $0.8\%$ as we apply larger bias $V_{dc}$. Hence, we think that $\Lambda^{out}/ \Gamma^{*}$ serves as a sensitive quantitative measure of the back-action strength.

\begin{table}[htb] 
\caption{Simulation results at various back-action condition} 
\label{Table1}
\begin{center}
\begin{tabular}{|c|c|c|c|c|c|c|}
\hline                  
$V_{T}$ & $V_{dc}$ & $\Lambda^{out}$ & $\Gamma^{*}$ & $\Lambda^{out}/\Gamma^{*}$ & $T$ & $E_{cut}$ \\
\hline       
$-1.40V$ & $0.5mV$ & $15.0Hz$ & $1.5kHz$ & $1.0\%$ & $0.32K$ & $3.75meV$ \\
\hline
$-1.46V$ & $0.5mV$ & $5.0Hz$ & $1.6kHz$ & $0.31\%$ & $0.32K$ & $3.40meV$ \\
\hline   
$-1.50V$ & $0.5mV$ & $0.5Hz$ & $2.5kHz$ & $0.026\%$ & $0.24K$ & $2.82meV$ \\
\hline       
$-1.50V$ & $1.5mV$ & $4.0Hz$ & $2.4kHz$ & $0.16\%$ & $0.38K$ & $2.85meV$ \\
\hline 
$-1.50V$ & $2.5mV$ & $20.0Hz$ & $2.5kHz$ & $0.8\%$ & $0.45K$ & $2.90meV$ \\
\hline
\end{tabular}
\end{center}
\end{table}

Also shown in Table I is that the electron temperature keeps warming up with $V_{dc}$, while only slightly increases with $V_{T}$. When $V_{T}$ goes from $-1.46V$ to $-1.40V$, the electron temperature nearly has no change although $\Lambda^{out}/ \Gamma^{*}$ increases by three times. This difference may indicate different back-action mechanisms. As we know, the inelastic back-action could be either direct or indirect. The QPC current heats up the electron bath and emits high frequency quasi-particles such as acoustic phonons \cite{Ludwig-Backaction-QPC-Current}. The partial re-absorption of these particles by the QD electrons causes indirect back-action. At large $V_{dc}$, the QPC current heats up the electron bath substantially and thus the indirect back-action could dominate. On the contrary, increasing the gap opening doesn't cause severe heating. Instead it increases the direct QD-QPC Coulomb coupling strength efficiently. In this case the direct back-action should prevail. In general, both the indirect and direct back-action should exist.

Another difference is in the cut-off energy $E_{cut}$, beyond where the saturation effect suddenly relieves. Fig. \ref{Figure3} (a) - (b) show $R_{0/1}$ in a large energy scale for varying $V_{dc}$ or $V_{T}$. We determined $E_{cut}$ as where $R_{0/1}$ drops to $10^{-6}$ since we only take $10^{6}$ data points. The results are listed in the last column of Table I. $E_{cut}$ shows very small change (between $2.82meV$ and $2.90meV$) with $V_{dc}$, and increases dramatically (from $2.82meV$ to $3.75meV$) for increasing $V_{T}$. In both cases, the observed $E_{cut}$ is much larger than $|eV_{dc}|$. Since $|eV_{dc}|$ is the maximum single phonon energy the QPC emits \cite{Kouwenhoven-Backaction-Theory, Ensslin-Backaction-eVqpc}, in our system there must involve multi-phonon absorption process \cite{Ludwig-Backaction-DQD}. And the independence on $|V_{dc}|$ indicates that $E_{cut}$ is not limited by the number of phonons to absorb. Instead, from its increase with $V_{T}$, we conclude that $E_{cut}$ has to do with the QD-QPC coupling strength. We found that $V_{T}$ effectively and exclusively controls the QD-QPC coupling. Fig. \ref{Figure3} (c) shows that with less negative $V_{T}$ the RTS amplitude, an indicator of the coupling strength,  increases quickly. On the contrary, the other gate voltages such as $V_{LB}$ changes neither the coupling strength nor the cut-off energy, as shown in Fig. \ref{Figure3} (d). Thus we suspect that $E_{cut}$ depends on $V_{T}$ through the QD-QPC coupling strength. However, at this moment we can not analytically determine this dependence. It may need further theoretical models, especially which incorporate multi-phonon process, to explicitly determine the cut-off energy.

In conclusion, we found that the non-equilibrium effect in QPC charge counting statistics is a benchmark of its back-action. We gave a quantitative measure of the back-action strength. This paved a way for further study, such as the back-action driven spin excitations \cite{Ming-Backaction-SpinST}.

This work was supported by the NFRP 2011CBA00200 and 2011CB921200, NNSF 10934006, 11074243, 10874163, 10804104, 60921091.

\end{document}